\documentclass{elsart}
\usepackage{epsfig}
\def\etal{{\rm et al.}}
\begin{document}
\hspace*{\fill}\large HEPSY 99-1\linebreak
\hspace*{\fill}\large Jan. 1999~\linebreak
\vspace*{0.5in}
\centerline{~~~}

\begin{frontmatter}

\title{Physics Results from RICH Detectors}

\author{Sheldon Stone$^1$}
\address{Syracuse University,
Syracuse, NY 13244-1130}
\thanks{Supported by the National Science Foundation}

\begin{abstract}
RICH detectors have become extraordinarily useful. Results include measurement
 of solar neutrino rates, evidence for neutrino oscillations, measurement of
 TeV $\gamma$-rays from gravitational sources, properties of QCD, charm
 production and decay, and measurement of the CKM matrix elements $V_{cs}$, 
$V_{cb}$ and $V_{ub}$. A new value $|V_{ub}/V_{cb}|=0.087\pm 0.012$ is 
determined.

\end{abstract}
\end{frontmatter}
\vspace{2cm}
\normalsize
\dotfill

{\em Invited talk at ``The 3rd International Workshop on Ring Imaging Cherenkov
Detectors," a research workshop of the Israel Science Foundation, Ein-Gedi,
Dead-Sea, Israel, Nov. 15-20, 1998}
\newpage


\section{Introduction}

This paper describes some physics results obtained by experiments using
Ring Imaging Cherenkov Detectors. Currently, the major physics issues are:
\begin{itemize}
\item{} What is the origin of mass? Another way of phrasing this question is to 
state that the Higgs boson must be found. This is a crucial reason for 
building the LHC machine. In this case, the RICH detectors have not yet been 
shown to have any relevance.
\item{} Do neutrinos have mass? If so can neutrino mixing be explained in the
 context of the Standard Model?
\item{} Does Quantum Chromo-Dynamics explain the strong interactions?
\item{} How to unify gravity with the other interactions?
\item{} Does the Standard Model explain quark mixing
via the CKM matrix? We need to measure the CKM elements and
CP violating angles. 
\end{itemize}

The ${\it raison~d'etre}$ for RICH detectors is to answer these
and other questions. The measurements reported here reflect
my own view and may be incomplete.

RICH detectors can be viewed as being used for two distinct
functions. One is to detect the presence of charged particles and
the other is to identify the kinds of particles that have been
detected by other devices.

\section{QCD Results}
One of the first RICH detectors was used in Fermilab experiment E605. They
measured particle spectra produced in 800 GeV/c proton interactions with 
Be and W nuclei. The detector used a He gas radiator and a He-TEA photon
detector. On average only 3 photo-electrons per track were detected. Yet 
results were produced \cite{E605}. 

\begin{figure}[hbt]
\centerline{\epsfig{figure=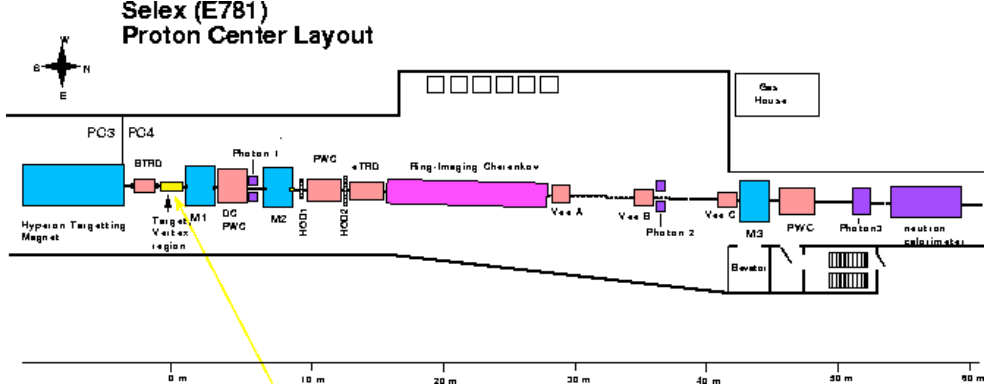,width=2in,angle=270}}
\caption{\label{selex_det} Diagram of the Selex detector. The dashed arrow
points to the vertex and target region, that contains the silicon strip
modules.}
\end{figure}

More sophisticated modern versions of fixed target spectrometers are 
the CERN OMEGA spectrometer and SELEX at Fermilab \cite{Selex}. A diagram of 
SELEX is
shown in Fig.~\ref{selex_det}. Precision silicon strip detectors are 
interspersed
with the target. A gas RICH detector with phototube readout is used to
indentify charged hadrons. 

\begin{figure}[hbt]
\centerline{\epsfig{figure=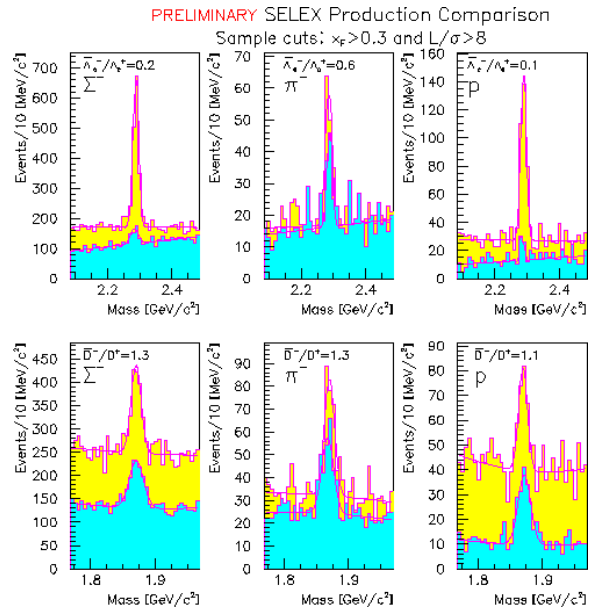,width=3in}}
\caption{\label{selex_charm} Invariant mass plots for $\Lambda_c^+$ and $D^+$
and their antiparticles for different incidents beams. The measured
production ratios are shown on the figure.}
\end{figure}

SELEX will study the decay of charmed baryons. They have already produced
results on the production ratios of charmed mesons and baryons from different
beams \cite{selex_res}, shown in Fig.~\ref{selex_charm}. These can be compared 
to 
theoretical
models of charm production.

\begin{figure}[hbt]
\vspace{-.3cm}
\centerline{\epsfig{figure=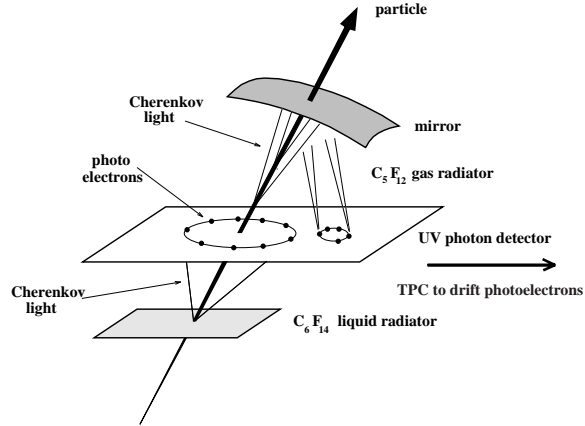,width=3in}}
\vspace{-.2cm}
\caption{\label{delphi_rich} A sketch of the DELPHI RICH and SLD CRID systems.}
\end{figure}

\begin{figure}[hbt]
\centerline{\epsfig{figure=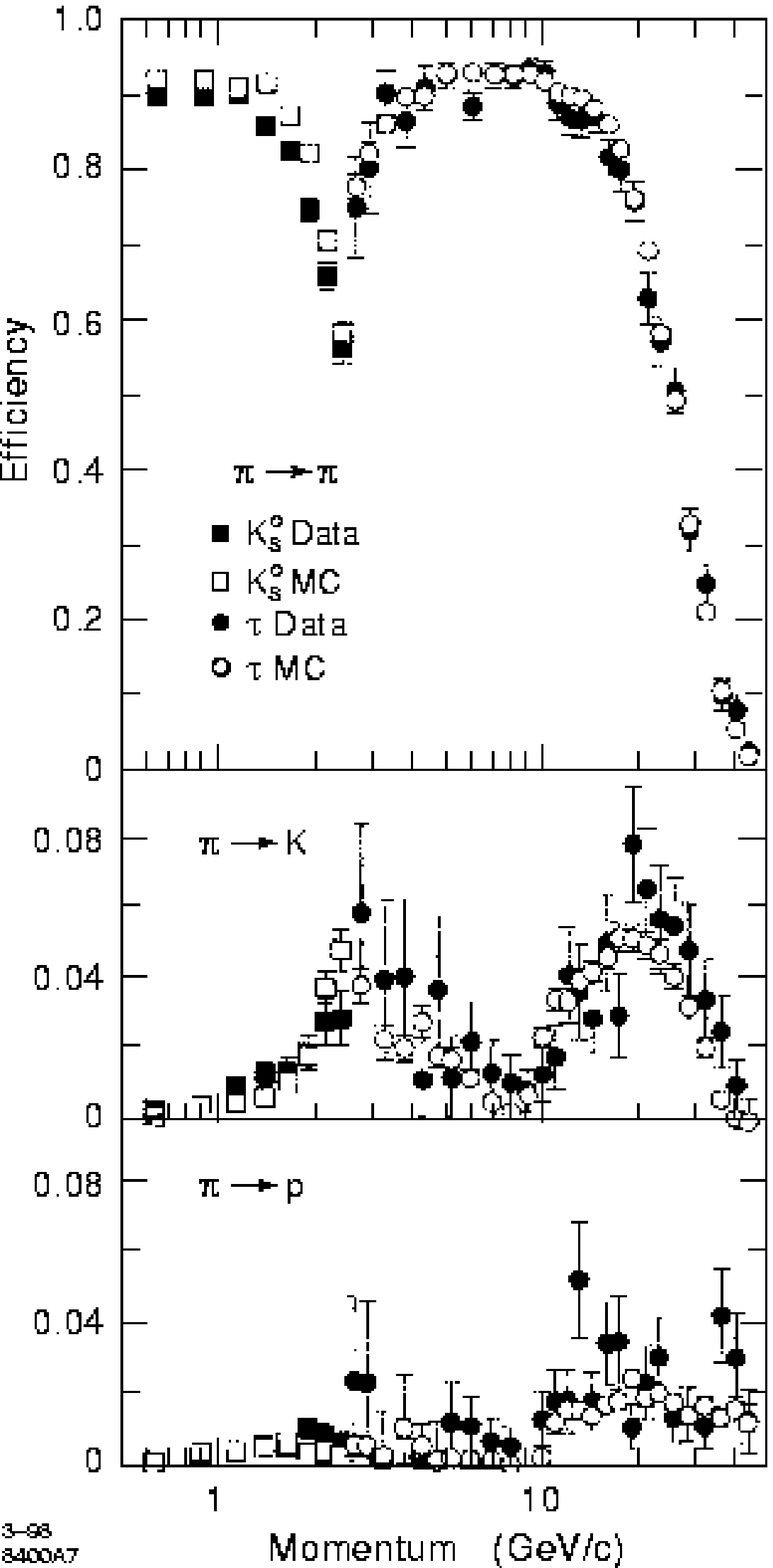,width=1.9in}\hspace{5mm}
\epsfig{figure=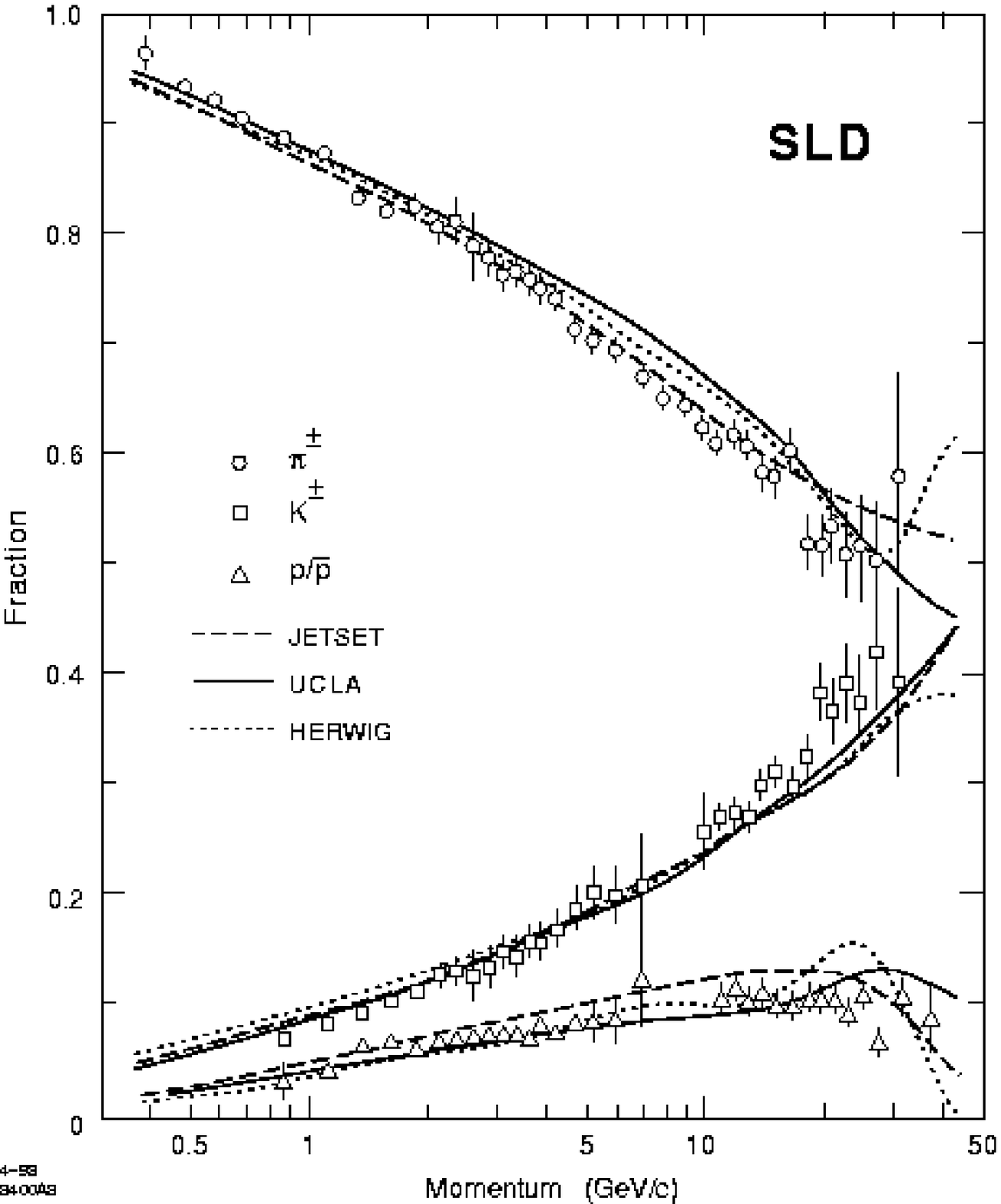,width=2.8in}}
\caption{\label{crid_eff}(Left) Efficiencies for tracks from $K_s^o$ (squares)
and $\tau^{\pm}$ (circles) decays to be identified as each hadron species
in the SLAC CRID. The solid symbols show the data and the open symbols the
simulation. (Right) Comparison of measured charged hadron fractions
(symbols) from SLD compared with the predictions of various fragmentation
models.}
\end{figure}

Two machines, LEP and SLC produce $Z^o$ bosons in $e^+e^-$ collisions.
Physics results have been obtained by two experiments using quite similar RICH 
detectors,
DELPHI \cite{DELPHI} at LEP and SLD \cite{SLD} at SLC. A primitive
sketch of the RICH systems is shown in Fig.~\ref{delphi_rich}.
These systems use both liquid C$_6$F$_{14}$ and gaseous fluorine radiators.
The Cherenkov photons are converted in TMAE and drifted using a TPC to proportional
wires. The performance of the SLD CRID has been characterized in terms of efficiency
and rejection for a particular set of analysis criterion. Shown in 
Fig.~\ref{crid_eff}(left) are efficiencies for the wanted hadron and the 
efficiency
for the unwanted species (really rejection).\footnote{In general, the
efficiency versus rejection is a function of the analysis cuts and can be
chosen differently to optimize each study.}

The $Z^o$ decays into quark-antiquark pairs or lepton-antilepton pairs.
In the case of the quarks, the energy ends up in hadrons and the momentum
spectrum of the hadron species can be compared with models of this
fragmentation.
The particle spectrum from $Z^o$ decays has been measured by both
DELPHI and SLD. The SLD spectrum is shown in Fig.~\ref{crid_eff}(right) and
compares well with the different models \cite{SLD_res}.

QCD predicts the presence of a new state of matter, a so called ``quark-gluon
plasma," that may be produced in the collisions of heavy nuclei. Evidence could
come dileptons produced in elementary collisions such as $q\overline{q}\to
\ell^+\ell^-$. Electron pairs are typically searched for. The can also come
from more mundane sources, such as vector meson decays or Dalitz decays.

\begin{figure}[hbt]
\vspace{-.2cm}
\centerline{\epsfig{figure=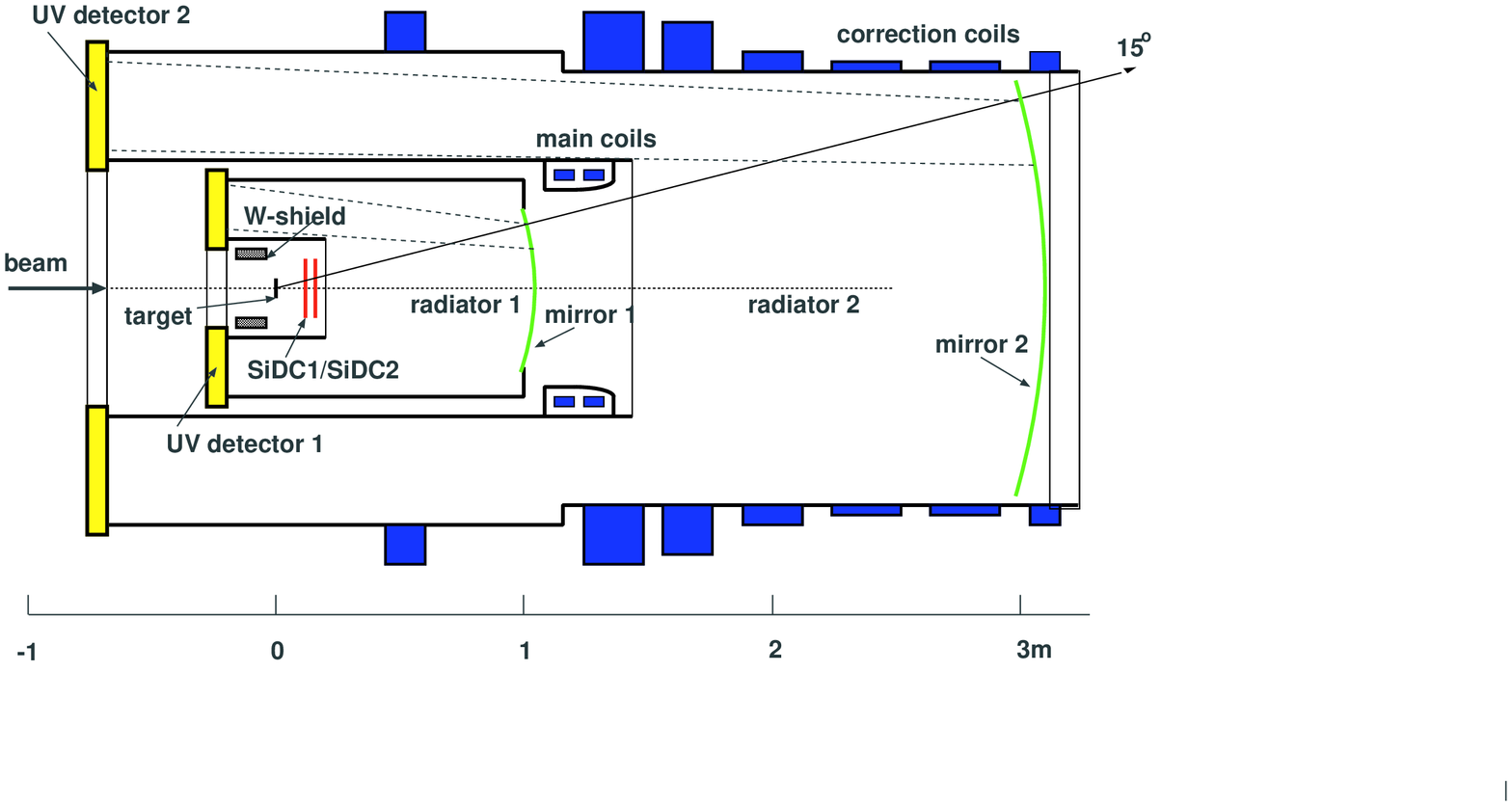,width=3in}\hspace{1mm}
\epsfig{figure=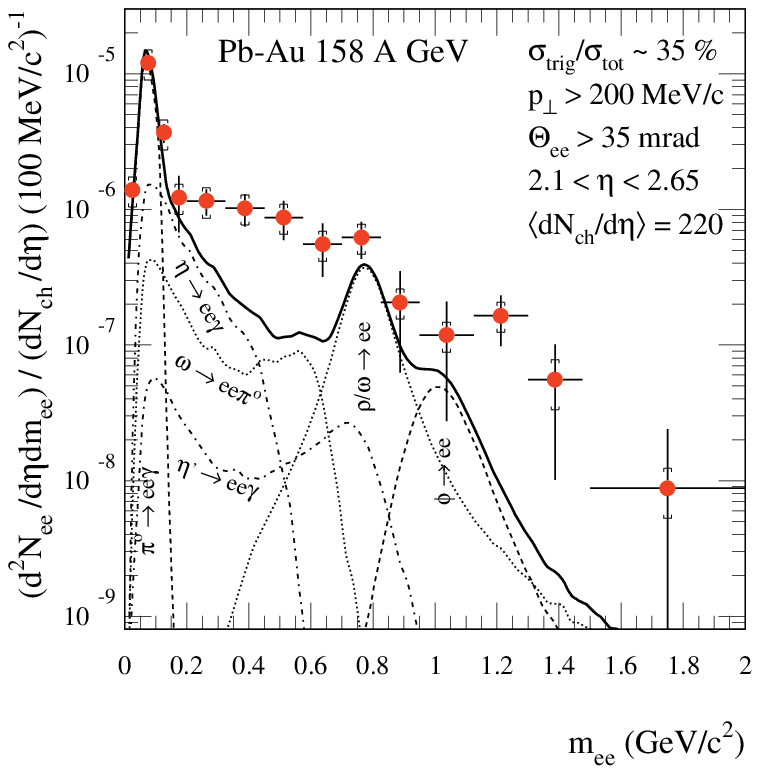,width=2.5in}}
\vspace{-.25cm}
\caption{\label{ceres}(Left) Sketch of the CERES detector. (Right)
The cross-section for $e^+e^-$ pairs as a function of dilepton mass.
The yield from expected sources is also shown. The statistical errors (bars)
and the systematic errors (brackets) are plotted independently of each
other.}
\end{figure}

The CERES experiment has searched for $e^+e^-$ pairs in heavy ion collisions at
the CERN SPS.
A sketch of the detector is shown in Fig.~\ref{ceres}(left). They have two RICH 
detectors using a CH$_4$ gas radiator and UV (TMAE based) photon detectors,
which are placed outside the region of the collision products \cite{CERES}. 
Also, 
heavier
particles, such as pions, tend not to radiate, so this type of system is
sometimes called ``hadron blind."

Fig.~\ref{ceres}(right) shows there results for Pb-Au collisions 
\cite{CERES_res}. There is an
excess of $e^+e^-$ pairs above the estimated background sources. Many new
experiments are being built to explore this phenomena including HADES at
GSI Darmstadt, and RHIC experiments at Brookhaven.

\section{Water Cherenkov Detectors}
\subsection{Introduction}
The quest for a theory to unify strong and electroweak interactions led to
models based on higher symmetry groups which predicted or at least allowed the
possibility of proton decay. The IMB and Kamiokande detectors were large tanks
of pure water placed deep underground to minimize the cosmic ray background.
The water provided the protons that might decay and also the means of
detection
using the Cherenkov light produced by the decay products. 
Kamiokande was greatly enlarged to Super-Kamiokande \cite{SuperK}. 

\begin{figure}[hbt]
\centerline{\epsfig{figure=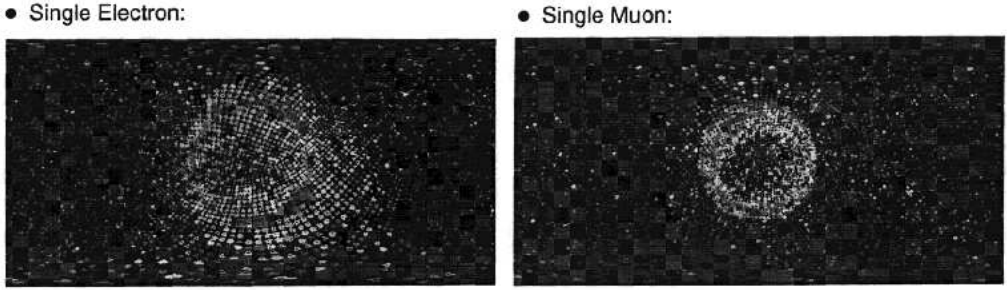,width=5.5in}}
\caption{\label{skmu}Examples of the detector response of an electron and a muon 
in
Super-Kamiokande.}
\end{figure}

All these
detectors have many large diameter photomultiplier tubes which gather the 
unfocused light (proximity focused in Tom Ypsilantis terminology). 
They can detect charged leptons and hadrons and also photons that convert
to pairs in the water. They can discriminate between electrons and muons
because the electrons scatter more. Fig~\ref{skmu} shows the phototubes hit
by an electron and muon. The width of the Cherenkov ring is much smaller in 
the muon case.

These detectors can also detect $\nu_e$ from the sun and from
supernovas. The latter capability was impressively demonstrated by detecting the
explosion of supernova 1987A \cite{supernova}. They also can detect neutrinos 
produced in the
atmosphere from cosmic ray interactions.

\subsection{The Atmospheric Neutrino Anomaly}
\begin{figure}[hbt]
\centerline{\epsfig{figure=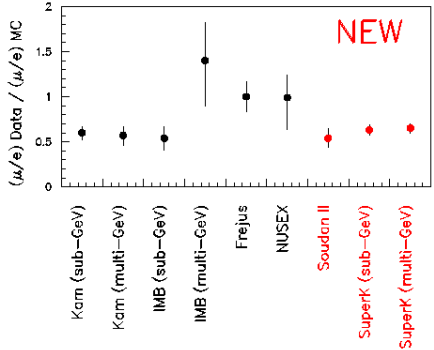,width=3in}}
\caption{\label{sk_atm}Summary of measurements of the atmospheric
neutrino ratio of measured muon/electron neutrino rates divided by the
expectation as given by Monte Carlo prediction. 
$(\nu_\mu/\nu_e)_{data}/(\nu_\mu/\nu_e)_{MC}$ for 
atmospheric experiments. The three newest results are listed at the right (from 
ref. \cite{conrad}).}
\end{figure}

Cosmic rays, mostly protons, hit the atmosphere and interact at altitudes
of ten to twenty kilometers above the Earth's surface. Naively, we expect
two muon-neutrinos, $\nu_{\mu}$, for every $\nu_e$, because of the
dominant decay chains
$\pi^+(K^{+})\to \mu^+\nu_{\mu}$; $\mu^+\to e^+\nu_e\bar{\nu_{\mu}}$ (and
similarly for $\pi^-(K^-))$. IMB first observed that the ratio was about half of
what was expected \cite{IMB_res}. The data from several experiments are shown in
Fig.~\ref{sk_atm}.

Although some experiments did not see the anomaly, the precise data from
Kamiokande and Super-Kamiokande \cite{mcgrew} clearly show that IMB was correct. 
The question
remained, however, was there something wrong in the expectation or was the
effect coming from a loss of $\nu_{\mu}$ due to neutrino oscillations?

In neutrino oscillations, different species can metamorphose
into other ones. We have three
known species of neutrinos, $\nu_e$, $\nu_{\mu}$ and $\nu_{\tau}$. If $\nu 's$
have mass they can transform from one to another. Specifically, the probability
that a neutrino, $\nu_i$, not oscillate is given by
\begin{equation}
P(\nu_i\to\nu_i)=1-\sin^2(2\theta) \sin^2\left(1.27\Delta m^2{L\over 
E_{\nu}}\right),
\end{equation}
where $\theta$ is the unknown mixing angle, $\Delta m$ is the unknown mass
difference, $L$ is the distance from the production point and $E_{\nu}$ the 
energy.

Cosmic rays hitting the upper atmosphere are distributed uniformly. The 
experiment is at a fixed point close to the Earth's surface.  If a
neutrino comes from directly overhead the cosine of its  angle is defined as
+1, while those coming through the earth have a cosine of -1. Thus $\nu 's$
have different path lengths to get to the detector; since the $\nu$
energy spectrum is the same from any direction, we can vary
the ratio $L/E_{\nu}$ by looking at different angles. These $\nu 's$ range
from below 1 GeV to many GeV in energy. (Super-Kamiokande defines sub-GeV to
be $<$1.3 GeV, and multi-GeV to be larger.)

\begin{figure}[hbt]
\vspace{-.2cm}
\begin{center}
\epsfig{figure=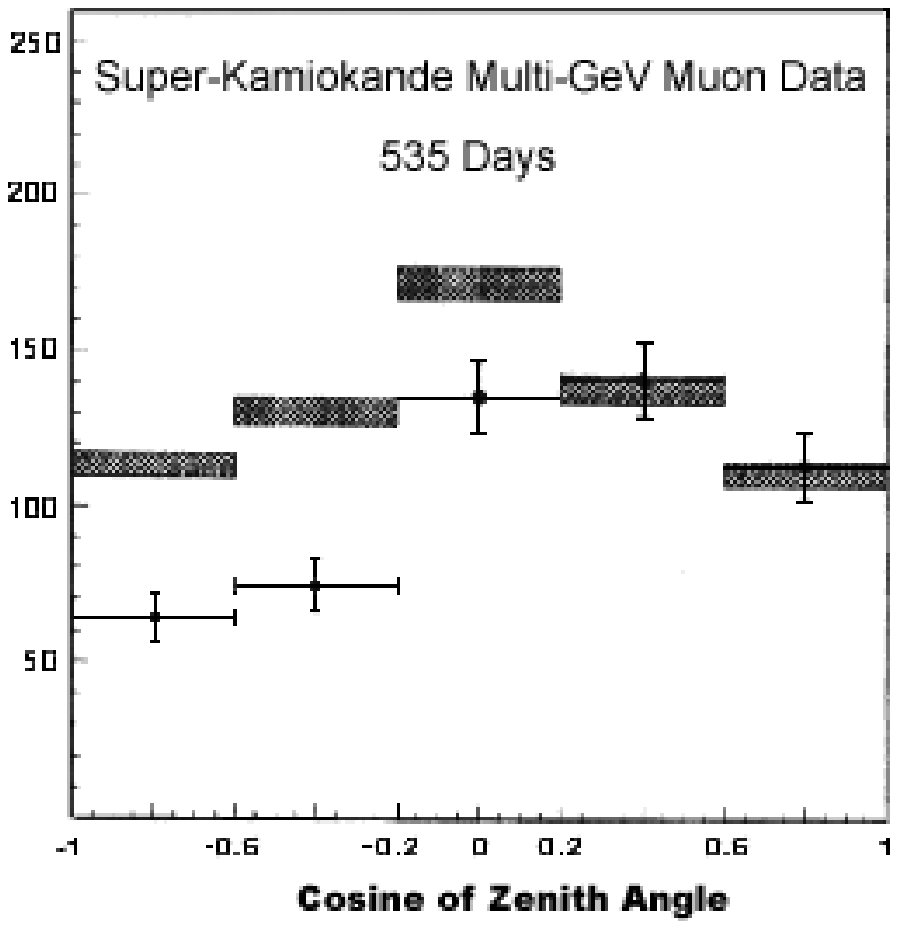,width=2.5in}\vspace{-3mm}
\epsfig{figure=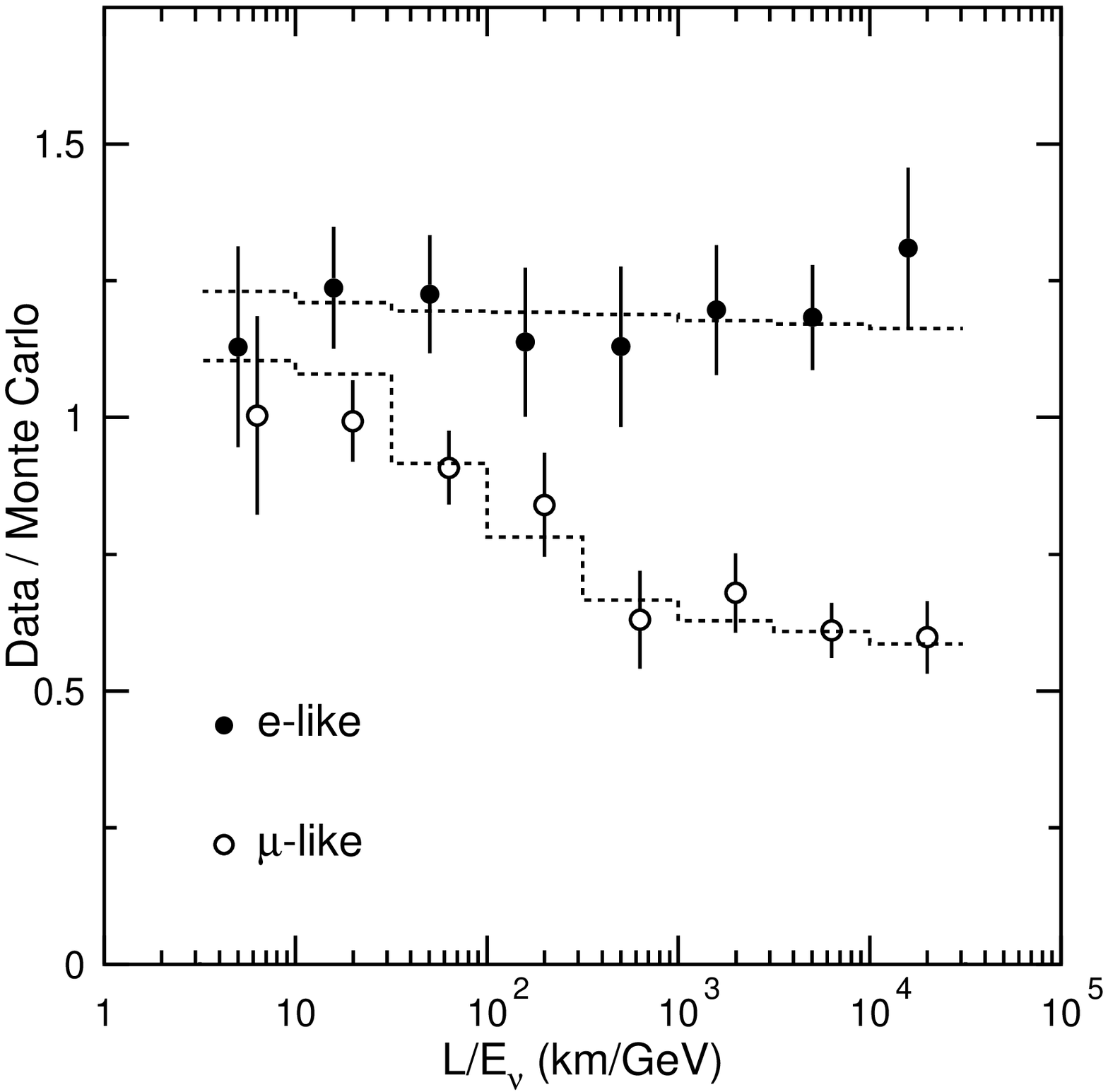,width=2.6in}
\end{center}
\caption{\label{sk_atmall}(Left) Zenith angle distributions of $\nu_{\mu}$
 multi-GeV events. Upward-going particles have $\cos
\theta < 0$ and downward-going particles have $\cos \theta > 0$.
 The shaded region shows the Monte Carlo expectation for no
oscillations normalized to the data live-time with statistical
errors. (Right) The ratio of data to Monte Carlo  versus
reconstructed $L/E_\nu$ (fully contained events only) \cite{mcgrew}.}
\end{figure}

 Super-Kamiokande is a large enough
detector to get enough atmospheric neutrino events to be able to show angular
distributions. While $\nu_e$ behave as expected,
the distribution of $\nu_{\mu}$ do not. In Fig.~\ref{sk_atmall}(left) the zenith
angle distribution of $\nu_{\mu}$ is shown along with the expectations based on
simulation. There is a clear deficit of up-going neutrinos. On the right side
the $L/E_{\nu}$ distribution is plotted for both $\nu_e$ and $\nu_{\mu}$. The
evidence for $\nu_{\mu}$ oscillating into something is clear for 
$\nu_{\mu}$. The dashed lines show the
expected shape for $\nu_{\mu}\Leftrightarrow \nu_{\tau}$ using
$\Delta m^2=2.2\times10^{-3} $eV$^2$ and $\sin^2(2\theta)$=1.
The dashed lines for $\nu_e$ show the expected shape without oscillations.
The slight $L/E_\nu$ dependence for $e$-like events is
due to contamination (2-7\%) of $\nu_{\mu}$. 

If the $\nu_{\mu}$ are oscillating, they can change into $\nu_e$,
$\nu_{\tau}$ or some other species of yet unobserved neutrino. As the latter
possibility is less likely, let us ignore it here. We do not see
an increase in $\nu_e$, so its most likely that the $\nu_{\mu}$ are 
changing into $\nu_{\tau}$. Unfortunately, the $\nu_{\tau}$ are not energetic
enough to produce $\tau^-$ via a charged current interaction, most of the
time.

\subsection{The Solar Neutrino Anomaly}

\begin{figure}[hbt]
\centerline{\epsfig{figure=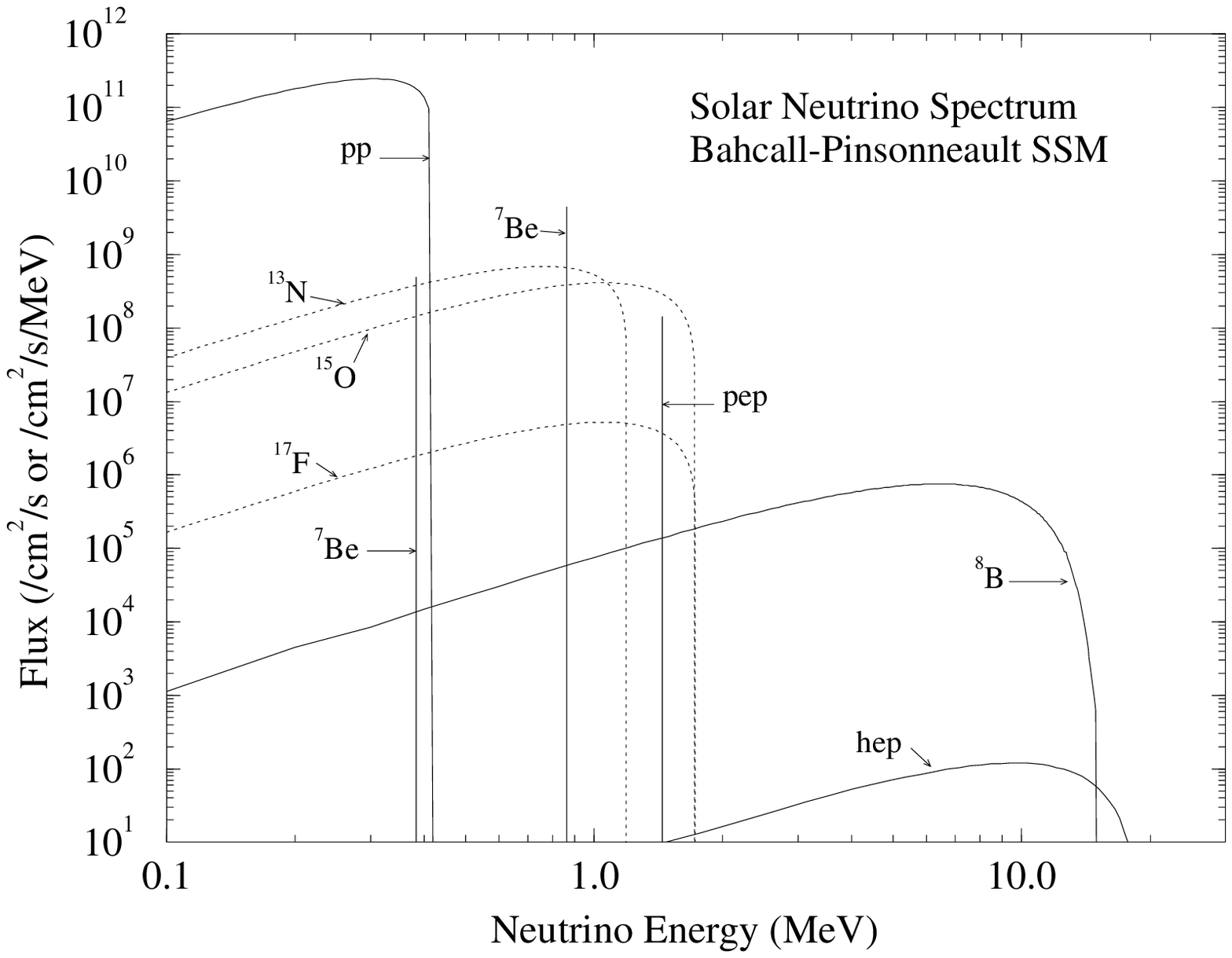,width=2.6in}
\epsfig{figure=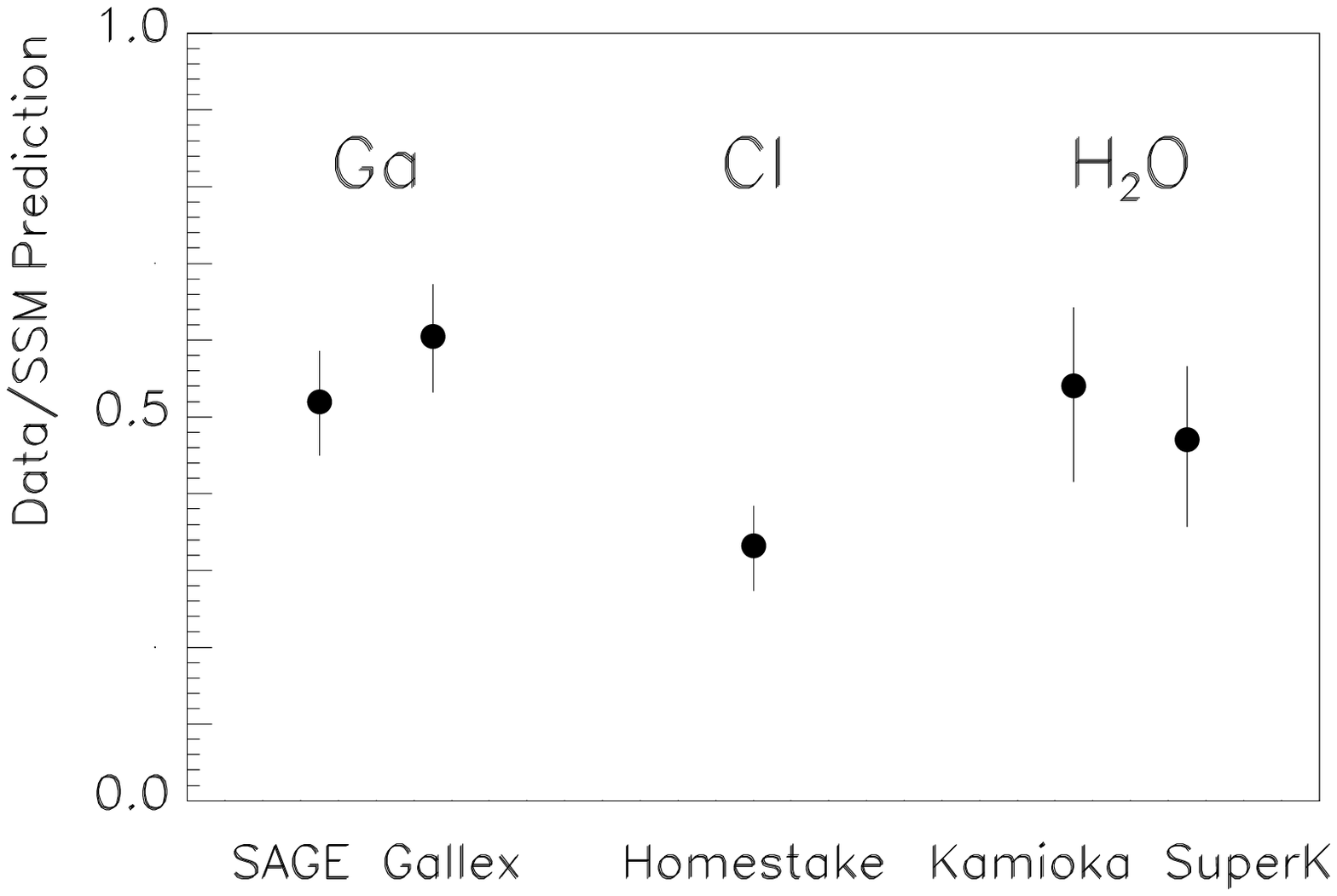,width=3.4in}}
\caption{\label{solar}(Left) The predicted solar neutrino energy spectrum.
The nuclear reactions are indicated. 
(Right) The measured solar neutrino flux from the different experiments 
divided by the expectations of the standard solar flux model. Gallium
experiments are sensitive above 0.23 MeV, Cl above 0.8 MeV, and H$_2$O above
6.5 MeV, from \cite{conrad}.}
\end{figure}

Thermonuclear reactions in the center of the sun are a prolific source
of $\nu_e$. Detecting them and comparing the rate with models describing
the process have gone on since 1964, starting with Ray Davis' pioneering 
experiment in
the Homestake mine \cite{davis}. Confirmation that the rate is about half what 
is expected
was made by the Gallium based experiments (Gallex and Sage) \cite{gallium}
 and the water
Cherenkov detectors \cite{sk_solar}. The expected flux, and the results are 
shown in 
Fig.~\ref{solar}.

\begin{figure}[hbt]
\centerline{\epsfig{figure=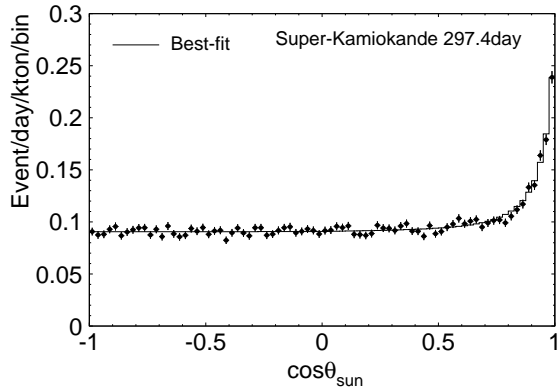,width=3in}}
\vspace{-.4cm}
\caption{\label{solar_dir} The solar neutrino spectrum
from Super-Kamiokande as a function of the angle with respect
to the sun.}
\end{figure}

Super-Kamiokande can also show that the detected $\nu_e$ come from the sun.
Fig~\ref{solar_dir} shows the angle of the produced electron with respect to 
the solar direction \cite{sk_solar}. The sharp peak is evident above the flat 
background.

One explanation of the $\nu_e$ deficit is that the sun is just at the right
distance from the earth that half have oscillated to other species. This is
called the ``Just So," solution \cite{justso}. Another explanation is there is a 
resonant
enhancement of oscillations due to the high electron density in the suns core;
this is called the MSW effect \cite{MSW}.

\subsection{Summary of Neutrino Oscillation Signals and Future Prospects}

\begin{figure}
\centerline{\epsfig{figure=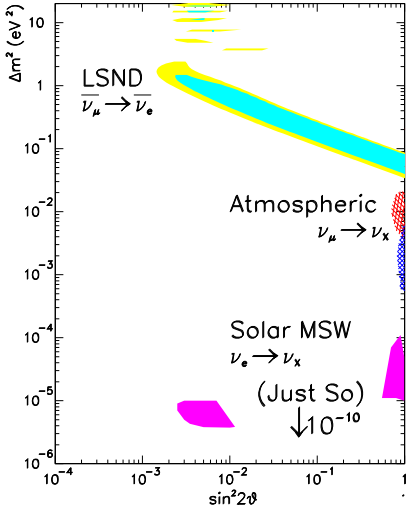,width=2.5in}}
\caption{\label{oscil}  Allowed regions for three indications of neutrino
oscillations. The top blob for Atmospheric is from Kamiokande while the bottom
is for Super-Kamiokande (from ref. \cite{conrad}).} 
\end{figure}

Possible signals for neutrino oscillations come from three sources, two
of which, atmospheric neutrinos and solar neutrinos have been described here.
A third source, the LSND experiment claims observation of a signal for
$\bar{\nu_{\mu}}\to \bar{\nu_e}$ \cite{LSND}. Putting these on the same $\Delta 
m^2 - \sin^2 (2\theta)$ plot (Fig.~\ref{oscil}) shows that it will be difficult 
to reconcile all three measurements.
 
In the near term future there will be several new experiments using Water
Cherenkov detectors that should reveal much about the mysteries just beginning
to be revealed. A beam of $\nu_{\mu}$ will be directed at Super-Kamiokande 
\cite{K2K}.
The SNO detector, a large D$_2$O detector will be able to detect the neutral
current neutrino interactions and thus be able to measure the total atmospheric
neutrino flux and the total solar flux as well as charged current interactions 
\cite{SNO}.
Borexino will start and be able to detect 0.86 MeV neutrinos from solar
Be$^7$ reactions at the level of 50 events/day \cite{borexino}.

\section{Measurement of Primary Cosmic Rays with Caprice}

Analysis of the atmospheric neutrino data relies on knowledge of the primary
cosmic ray flux. The Caprice experiment has taken data with a magnetic
spectrometer incorporating a RICH detector with a NaF radiator and photon
detector using a wire-proportional chamber filled with TMAE and ethane 
\cite{caprice_det}. They
also have tracking and an electromagnetic calorimeter.

\begin{figure}
\centerline{\epsfig{figure=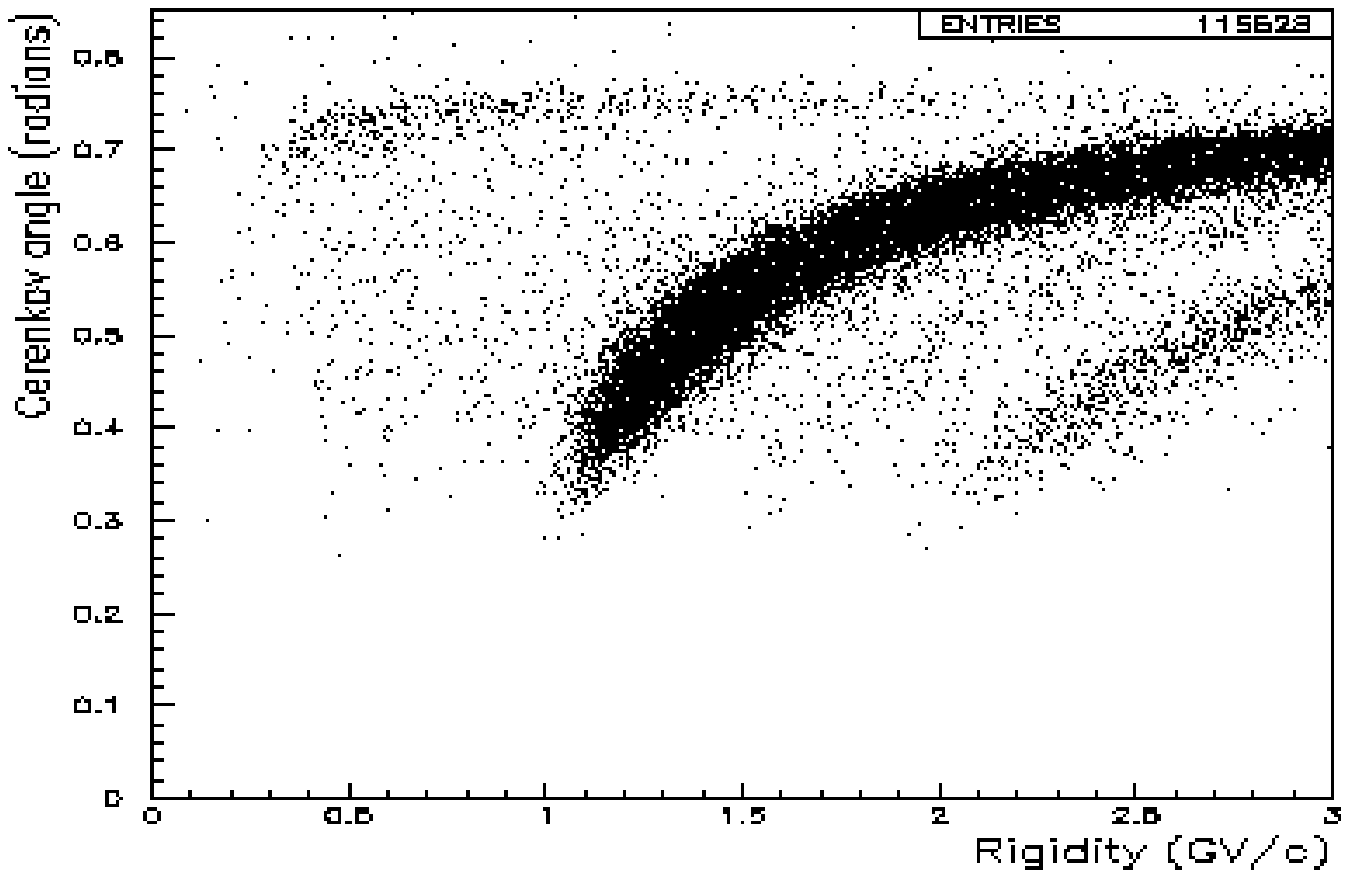,width=2.5in},
\epsfig{figure=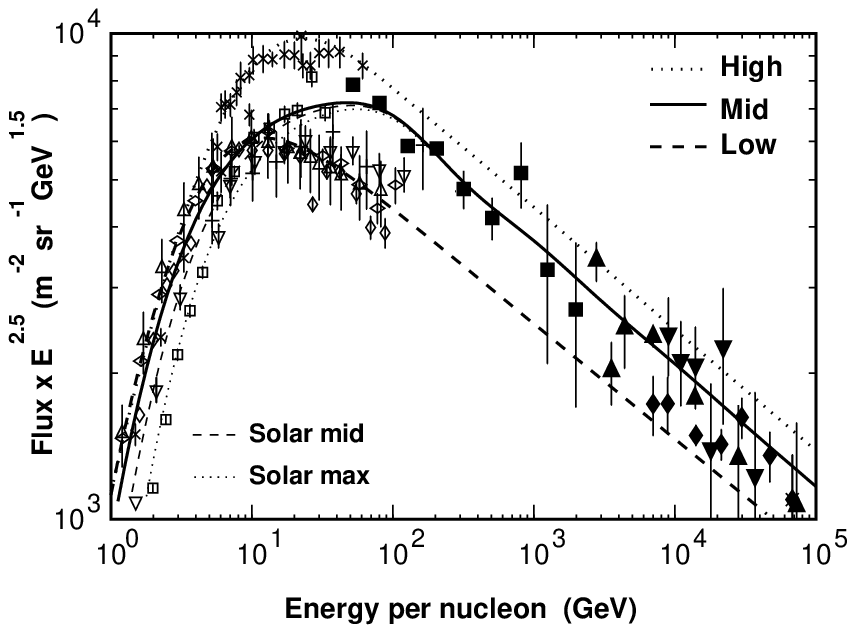,width=2.9in}}
\caption{\label{caprice}  (Left) Cherenkov angle as a function of rigidity, 
$R$, for positive particles in CAPRICE, where $R= (A/Q) (E^2 + 2M_oE)^{1/2}$,
 $M_o$=0.9315 GeV, $A$ is  atomic mass number, and $Q$ is charge. The bands,
top to bottom, correspond to $\mu^+$, $p$ and He nuclei. (Right)
Measured primary cosmic ray proton flux. The CAPRICE data are shown by open 
vertical
diamonds \cite{caprice_res}. Other data are described in ref. \cite{others}. 
The labels High, Mid and Low refer to the
ranges of fluxes used by HKKM \cite{others}.}
\end{figure}

CAPRICE has  measured the rates of many species at the top of the atmosphere.
The response to positive tracks is shown in Fig.~\ref{caprice}(left). 
Their primary cosmic ray (mostly proton)
flux measurement \cite{caprice_res} is compared with others in Fig.~\ref{caprice}(right). The data
that shows a larger rate is from older measurements. They have also
 measured the $\overline{p}/p$ rate to be $\sim 10^{-4}$, 
$\mu^+/\mu^-$=1.64$\pm$0.08 and $e^+/e^-$ to be $\sim 0.1$

Calculations of the primary cosmic ray flux and their interactions
with air, producing secondary mesons that then decay into leptons and
neutrinos are important in evaluating the total neutrino yield and less
consequential in evaluating the angular distribution of the $\nu_{\mu}/\nu_e$
ratio.
Super-Kamiokande uses several calculations of primary cosmic ray flux
\cite{hkkm}\cite{cosmic_flux}.
The HKKM \cite{hkkm} calculation of the primary cosmic ray
flux is shown as ``Mid" on Fig.~\ref{caprice}(right). Francke has argued at this 
meeting that the
HKKM calculation over estimates the CAPRICE data by about a factor of two. This
could lead to some reinterpretation of the atmospheric neutrino data and may
change the limits in the $\Delta m^2-\sin^2(2\theta)$ plane.

\section{Gravitational Physics with Cherenkov Detectors}
The Fly's Eye detector in Utah uses scintillation light to detect
ultra-relativistic cosmics rays \cite{flies_eye}. Cherenkov light produced in 
the 
atmosphere
from TeV $\gamma$-rays, after converting to $e^+e^-$ pairs can also be used.
HEGRA incorporates several Cherenkov light detectors with
scintillation detection of electromagnetic particles, and muon detection
over an area of 200x200 m$^2$ \cite{hegra_det}.\footnote{For a descripton
of other Cherenkov air shower experiments, see E. Lorentz in these
proceedings.}

\begin{figure}
\centerline{\epsfig{figure=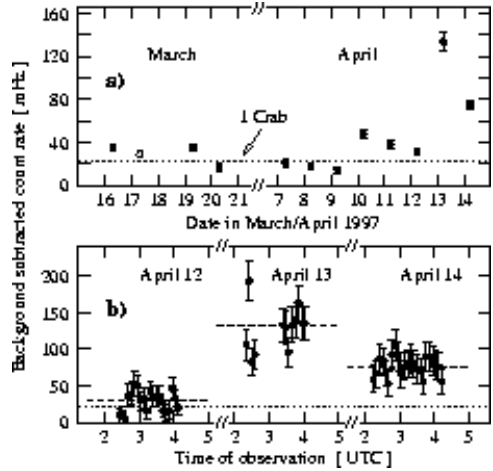,width=2.5in}}
\caption{\label{hegra} Detection rate of Mkn 501 on a night by night
basis (a) for the whole data sert and in 5 min. intervals (b) for the
last 3 nights. The dashed lines give the average per night, the dotted
line gives the Crab rate.}
\end{figure}

One example of what they have observed is shown in Fig.~\ref{hegra}.
They measured a high flux of TeV $\gamma$ rays from the source Mkn501.
The energy spectrum follows a power law from 1-10 TeV. Furthermore, the source
intensity varies from half that of the Crab Nebula to six times in very small
time intervals. This phenomena has not been explained, but clearly these
measurements will lead to an increased understanding of gravitational physics
phenomenology \cite{hegra_res}.

\section{Measurment Of The Cabibbo-Kobayshi-Maskawa Matrix}
\subsection{Introduction}
In the ``Standard Model" of Electroweak interactions,
the gauge bosons, $W^{\pm}$, $\gamma$ and $Z^o$ couple to  
mixtures of the physical $d,~ s$ and $b$ quark states. This mixing is described
by the Cabibbo-Kobayashi-Maskawa (CKM) matrix  \cite{ckm}.
\begin{equation}
V_{CKM} =\left(\begin{array}{ccc} 
V_{ud} &  V_{us} & V_{ub} \\
V_{cd} &  V_{cs} & V_{cb} \\
V_{td} &  V_{ts} & V_{tb}  \end{array}\right),\end{equation}
where the subscripts refer to the quarks.
In the Wolfenstein 
approximation\footnote{In higher order other terms have an imaginary part; in
particular the $V_{cd}$ term becomes $-\lambda-A^2\lambda^5(\rho+i\eta)$, which
is important for CP violation in $K^o_L$ decay.} 
~the matrix is written in terms of the parameters $\lambda$, $A$, $\rho$ and 
$\eta$
as \cite{wolf}
\begin{equation}
V_{CKM} = \left(\begin{array}{ccc} 
1-\lambda^2/2 &  \lambda & A\lambda^3(\rho-i\eta) \\
-\lambda &  1-\lambda^2/2 & A\lambda^2 \\
A\lambda^3(1-\rho-i\eta) &  -A\lambda^2& 1  
\end{array}\right)
\end{equation}
These parameters are {\bf fundamental constants} of nature that 
need to be determined experimentally, as is required for other
fundamental constant such as $\alpha$ or $G$. 

The constants $\lambda$ and $A$ are determined from charged-current weak 
decays, and are 0.22 and $\sim$0.8, respectively \cite{virgin}. The $\eta$ term
gives rise to a complex phase in the matrix which
allows CP violation in weak interactions. A similar matrix may exist for
$\nu 's$, if they have mass. It would explain $\nu$ mixing and predict
the possibility of CP violation in $\nu$ interactions.

\begin{figure}
\centerline{\epsfig{figure=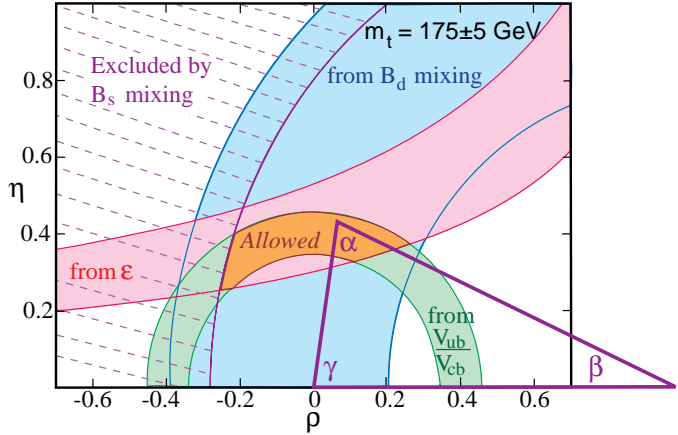,width=3.5in}}
\caption{\label{ckm_tri}The regions in $\rho-\eta$ space (shaded) consistent
with measurements of CP violation in $K_L^o$ decay ($\epsilon$), 
$V_{ub}/V_{cb}$ in semileptonic $B$ decay, $B_d^o$ mixing, and the excluded
region from
limits on $B_s^o$ mixing. The allowed region is defined by the overlap of
the 3 permitted areas, and is where the apex of the  CKM triangle  sits. The
bands represent $\pm 1\sigma$ errors. The error on the $B_d$ mixing band is
dominated by the parameter $f_B$. Here the range  is taken as 
$240> f_B > 160$ MeV. $|V_{ub}/V_{cb}|$ is taken as 0.087$\pm$0.012 
(see section 6.3). }
\end{figure}

Although $\rho$ and $\eta$ have not been determined, there are measurements
that provide constraints on their value. These include CP violation in $K_L^o$
decay, characterized by the value $\epsilon$, $B^o\Leftrightarrow\overline{B}^o$ 
mixing
and $V_{ub}/V_{cb}$. My interpretation of the current status of these 
measurements is shown in Fig.~\ref{ckm_tri}. Here I have used only $\pm 1\sigma$
errors, with the caveat that the dominant errors in all three cases are
caused by estimates of uncertainties on theoretical parameters.

Study of these processes has suffered from lack of particle identification 
capabilities.
CLEO and ARGUS had time-of-flight scintillators and
dE/dx measurements from the tracking chamber \cite{cleoii_arg}. However,
they are blind to $K/\pi$ separation between $\sim 1-2.2$ GeV/c
with poor separation above 2.2 GeV/c. Yet CLEO was the first to measure
$|V_{ub}/V_{cb}|$ and both made the first relatively precise measurements of 
$|V_{cb}|$. Recently, the 
DELPHI RICH has brought some improvements, which point to future possibilities.

\subsection{Measurement Of $V_{cb}$ Using $B\to D^*\ell\nu$}
Currently, the most favored
technique is to measure the decay rate of $B\to D^{*}\ell^-\bar{\nu}$ at the
kinematic point where the $D^{*+}$ is at rest in the $B$ rest frame (this is 
often referred to as maximum $q^2$ or $\omega =1$). Here, the theoretical 
uncertainties are at a minimum.\footnote{Current estimates of the form-factor 
necessary to translate the decay rate measurement into a value are 0.91$\pm 
0.03$ \cite{formfactor}.}
There have been previous results using this technique using the decay sequence 
$D^{*+}\to \pi^+ D^o$; $D^o\to
K^-\pi^+$, or similar decays of the $D^{*o}$.

\begin{figure}
\centerline{\epsfig{figure=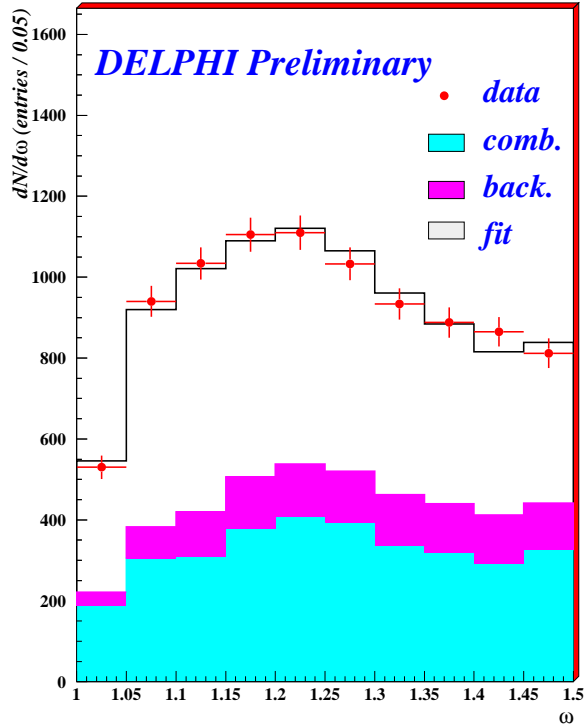,width=3.0in}}
\caption{\label{delphi_qsq}Fit to the $\omega$ distribution from DELPHI.
The solid points are the data, the light shaded area corresponds to
combinatorial background, the dark shaded area correspond to other background
and the horizontal line contains these components plus the signal.}
\end{figure}

In a recent analysis, DELPHI detects  only the slow $\pi^+$ from the $D^{*+}$ 
decay and does not 
explicitly reconstruct the $D^o$ decay \cite{delphi_pi}. The RICH helps in 
measuring missing 
energy in this complicated measurement. The distribution of events with 
respect to $\omega$ is shown in Fig.~\ref{delphi_qsq}. The dip in the data
near $\omega$ of one is an artifact caused by the drop in efficiency in 
detecting 
the slow $\pi^+$ at low momentum. Still, the decay rate can be ascertained.
Table~\ref{tab:Vcb} summaries determinations of $V_{cb}$. Here,
the first error on is statistical, the second systematic and the third, an estimate 
of the theoretical accuracy in predicting the form-factor at $\omega$ of 1. 
Currently, DELPHI has the smallest error, 
however, CLEO has only used 1/6 of their current data and will certainly improve 
on this measurement. The quoted average $|V_{cb}|=0.0381\pm 0.0021$
combines the averaged statistical and systematic errors with the theoretical 
error in quadrature and
takes into account the common systematic errors, such as the $D^*$ branching
ratios.

\begin{table}[th]
\begin{center}
\caption{Modern Determinations of $V_{cb}$ using 
$B\to D^*\ell^-\overline{\nu}$ decays
at $\omega = 1$ \label{tab:Vcb}}
\vspace*{2mm}
\begin{tabular}{lc}\hline\hline
Experiment & $V_{cb}$ $(\times 10^{-3})$\\\hline
DELPHI \cite{delphi_pi} & $41.2\pm 1.5 \pm 1.8 \pm 1.4$ \\
ALEPH \cite{aleph_pi} & $34.4\pm 1.6 \pm 2.3 \pm 1.4$ \\
OPAL \cite{opal_pi} & $36.0\pm 2.1 \pm 2.1 \pm 1.2$ \\
CLEO \cite{cleo_pi} & $39.4\pm 2.1 \pm 2.0 \pm 1.4$ \\\hline
Average & $38.1\pm 2.1$\\
 \hline\hline
\end{tabular}
\end{center}
\end{table}

\subsection{Measurement Of $V_{ub}$}
Another important CKM element that can be measured using semileptonic decays is
$V_{ub}$. The first measurement of $V_{ub}$ done by CLEO and subsequently
confirmed by ARGUS, used only leptons which were more energetic than those that
could come from $b\to c\ell^- \bar{\nu}$ decays \cite{first_vub}. These  
``endpoint leptons'' can occur, $b\to c$ background free, at the
$\Upsilon (4S)$ because the $B$'s are almost at rest. Unfortunately, there is
only  a small fraction of the $b\to u \ell^-\bar{\nu}$ lepton spectrum that
can be seen this way, leading to model dependent errors in extracting final 
values. 

\begin{figure}
\vspace{-9mm}
\centerline{\epsfig{figure=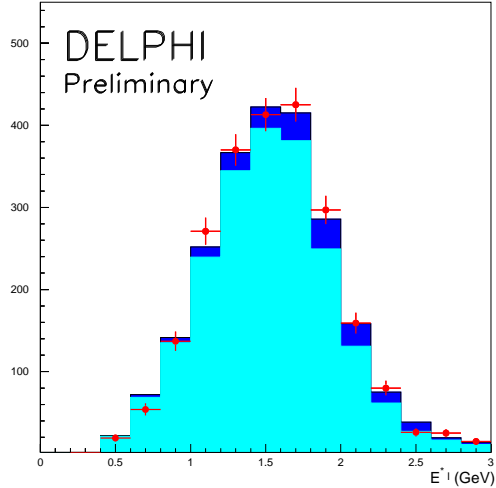,width=2.6in}}
\vspace{-4mm}
\caption{\label{delphi_vub}The lepton energy distribution in the $B$ rest
frame from DELPHI. The data have been enriched in $b \to u$ events, and the 
mass of the recoiling hadronic system is required to be below 1.6 GeV. The 
points indicate data, the light shaded region, the fitted background and the 
dark shaded region, the fitted $b \to u \ell \nu$ signal. The theoretical
models are described in ref. \cite{vub_thy}.}
\end{figure}

DELPHI tries to isolate a class of events where the hadron system associated
with the lepton is enriched in $b\to u$ and thus depleted in $b\to c$ 
\cite{delphi_vub}.     
They define a likelihood that hadron tracks come from $b$ decay by using a large 
number of variables including, vertex information, transverse momentum, not 
being a kaon. Then they require the hadronic mass to be less than 1.6 GeV, which 
greatly reduces $b\to c$, since a completely reconstructed $b\to c$ decay has a 
mass greater than that of the $D$ (1.83 GeV). They then examine the lepton 
energy distribution for this set of events, shown in Fig.~\ref{delphi_vub}.

DELPHI finds $\left|V_{ub}/V_{cb}\right|=0.104\pm 0.12 \pm 0.015 \pm 0.009$. The 
first error is statistical, the second systematic and the third the uncertainty 
quoted by Uraltsev on his model that allows the extraction of $|V_{ub}|$ from the 
measured branching ratio \cite{vub_thy_inc}. The systematic error is larger than 
the statistical error. This reflects the extensive modeling of the $b\to c$ 
decays. The theoretical error is estimated at 8\%. However, another calculation 
using the same type of model by Jin \cite{jin} gives a 14\% lower value, with a 
quoted error of 10\%. 

\begin{figure}
\centerline{\epsfig{figure=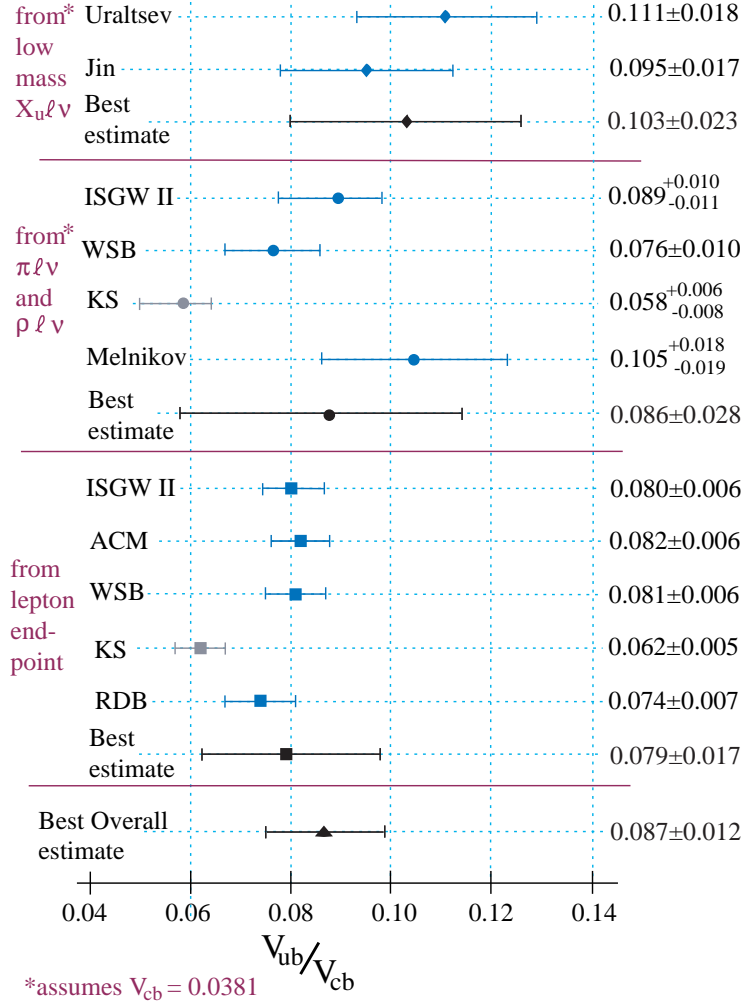,width=4.0in}}
\caption{\label{vub}Measurements of $|V_{ub}/V_{cb}|$ using different
techniques and theoretical models \cite{vub_thy}. (The KS model has been ruled 
out.) }
\end{figure}

ALEPH \cite{aleph_vub} and L3 \cite{L3_vub} have used techniques similar to 
DELPHI. I have averaged all three LEP results and shown them in 
Fig.~\ref{vub} without any theoretical error.  
My best estimate of $\left|V_{ub}/V_{cb}\right|$ 
using this technique includes a 14\% theoretical error added in quadrature
with a common systematic 
error of 14\%, since the Monte Carlo calculations at LEP are known to be 
strongly correlated. 

Also shown in Fig.~\ref{vub} are results from CLEO using the measured the decay 
rates for the exclusive final states $\pi\ell\nu$ and $\rho\ell\nu$ 
\cite{cleo_pirho}, and results from endpoint leptons, dominated 
by CLEO II \cite{cleo_vub}. From the exclusive results, the model of Korner and 
Schuler (KS) is ruled out by the measured ratio of $\rho/\pi$. This model deviated 
the most from the others used to get values of $|V_{ub}|$ from endpoint leptons. 
Thus the main use of the exclusive final states has been to restrict the 
models. The endpoint lepton results are statistically the most precise. 
Assigning a model dependent error is quite difficult. I somewhat arbitrarily 
have assigned a 14\% irreducible systematic error to these models and used the 
average among them to derive a value.
My best overall estimate is that $|V_{ub}/V_{cb}|=0.087\pm 0.012$.

\subsection{Measurement of $V_{cs}$}
Using the reaction $e^+e^-\to W^+ W^-$ at LEP II DELPHI has directly measured 
the CMK element $V_{cs}$ There are two ways that LEP II experiments have 
estimated this value. One way, is to measure the branching ratio of the $W's$
into hadrons and use the sum rule:
\begin{equation}
\left|V_{cs}\right|=
\sqrt{
{{3\Gamma_{\ell}}\over{\Gamma_h}}
{{{  Br}(W^{\pm}\to hadrons)}\over{(1-{ Br}\left(W^{\pm}\to 
hadrons      )\right)}}
-\sum_{ij\ne cs}\left|V_{ij}\right|^2},
\end{equation}
where $\Gamma_{\ell}$ and $\Gamma_{h}$, are the leptonic and hadronic widths of 
the $W^{\pm}$.

A better way, which does not rely on other measurements to measure the ratio 
directly of ${{{Br }(W\to \overline{c} s)}/ {{Br}(W\to hadrons)}}$.
For this DELPHI uses the RICH to identify kaons, which come not only from the 
$s$ quark but are also prolifically produced in $\overline{c}$ decays. Using this method 
DELPHI finds 
\begin{equation}
\left|V_{cs}\right|=1.01^{+0.12}_{-0.10}\pm 0.10,
\end{equation}
which is by far the best direct measurement of this quantity, and is consistent 
with the expected value of 0.95. This is important
because it checks the predictions of the CKM hypothesis.\footnote{The indirect 
method gives $\left|V_{cs}\right|=0.98\pm 0.07\pm 0.04$~~.}
\section{Conclusions}
It takes a Village to build a RICH, a village of knowledge cultivated at this 
conference. 
There are several different technologies: gas, H$_2$O, and crystal radiators 
coupled with gas, CsI, phototube photo-electron detectors, producing great 
results in neutrino physics, $b$ decays  and QCD. 
We expect even more interesting results soon!

\section{Acknowledgments}
I thank  Amos Breskin and Rachel Chechik for arranging an excellent meeting. 
My collegues Marina Artuso, Ray Mountain and Tomasz Skwarnicki helped on various 
aspects of this talk.

\end{document}